\documentclass[amsmath,amssymb,twocolumn,nofootinbib]{revtex4-1}
\usepackage[dvipsnames]{xcolor}
\usepackage{graphicx,epsfig,hyperref}
\hypersetup{unicode=true,colorlinks=true,linkcolor=NavyBlue,citecolor=NavyBlue,urlcolor=black}
\usepackage[normalem]{ulem}
\newcommand{\e}{\text{e}}
\begin{document}
\title{Bianchi cosmologies, magnetic fields and singularities}
\author{Roberto Casadio}
\email{roberto.casadio@bo.infn.it}
\affiliation{Dipartimento di Fisica e Astronomia, Universit\`{a} di Bologna,
via Irnerio 46, 40126 Bologna, Italy 
\\
I.N.F.N., Sezione di Bologna, I.S.~FLAG, viale B.~Pichat 6/2, 40127~Bologna, Italy}
\author{Alexander Kamenshchik}
\email{kamenshchik@bo.infn.it}
\affiliation{Dipartimento di Fisica e Astronomia, Universit\`{a} di Bologna,
via Irnerio 46, 40126 Bologna, Italy 
\\
I.N.F.N., Sezione di Bologna, I.S.~FLAG, viale B.~Pichat 6/2, 40127~Bologna, Italy}
\author{Panagiotis Mavrogiannis}
\email{pmavrogi@physics.auth.gr}
\affiliation{Section of Astrophysics, Astronomy and Mechanics, Department of Physics,
Aristotle University of Thessaloniki, Thessaloniki~54124, Greece}
\author{Polina Petriakova}
\email{polina.petriakova@bo.infn.it}
\affiliation{Dipartimento di Fisica e Astronomia, Universit\`{a} di Bologna,
via Irnerio 46, 40126 Bologna, Italy 
\\
I.N.F.N., Sezione di Bologna, I.S.~FLAG, viale B.~Pichat 6/2, 40127~Bologna, Italy}
\begin{abstract}
We study the effects of a spatially homogenous magnetic field in Bianchi-I cosmological models.
The cases of a pure magnetic field and two models with additional dust and a massless scalar field
(stiff matter) are also considered.
At the beginning of the cosmological evolution, {\em i.e.},~in the neighborhood of the singularity,
the universe is described by one of Kasner's solutions, and asymptotically by another Kasner
solution when the volume of the universe tends to infinity.
The transition law between these two Kasner regimes is established, and shown to coincide
with the analogous law for the empty Bianchi-II universe.
The universe filled with dust and a magnetic field undergoes the process of isotropization,
while the presence of a massless scalar field induces a modification of the relations between
Kasner indices in the two asymptotic regimes. In all of these cases, we analyze the approach
to the singularity in some detail and comment on the issue of the possible singularity crossing.
\end{abstract}
\maketitle
\section{Introduction}
\label{s:intro}
Almost all of modern cosmology is based on the spatially homogeneous
and isotropic Friedmann--Lema\^itre cosmological models.
Indeed, the Friedmann{--}Lema\^itre cosmology for an expanding or contracting spatially homogeneous and isotropic
universe is very successful in describing the global evolution of the universe
from inflation to the present epoch of cosmic acceleration.
Moreover, inhomogeneities in our universe are described by means of the theory of cosmological perturbations
on the Friedmann background, which explains the origin of large-scale structures in the contemporary universe
starting from quantum fluctuations in the very early universe.
\par
However, there are very interesting features in gravity and cosmology beyond the Friedmann models
and perturbations on such backgrounds. 
The study of spatially homogeneous but anisotropic models (see, e.g.,~Refs.~\cite{Land-Lif,Ryan,Bel-Hen})
originating in the work of Bianchi, published as early as in the year 1898~\cite{Bianchi}, presents great interest
from both the mathematical and physical points of view. Bianchi elaborated on the complete classification of
the three-dimensional homogeneous Riemannian spaces and three-dimensional Lie groups long before Einstein
put forward General Relativity in 1915.
Bianchi classification was then modernized, simplified, and applied to cosmology in the 50's and 60's~\cite{Held,Kras,Ellis}. 
\par
The simplest spatially homogeneous and anisotropic cosmology is given by the Bianchi-I model.
Its isometry group contains three generators (Killing vector fields), which commute and correspond to the metric
\begin{equation}\label{Bianchi-I}
ds^2 = dt^2 - a^2(t)\,dx^2 - b^2(t)\,dy^2 - c^2(t)\,dz^2
\ .
\end{equation}
One can see that Eq.~\eqref{Bianchi-I} reduces to a flat Friedmann metric in the limiting case when
the three scale factors $a(t)$, $b(t)$, and $c(t)$ coincide.  
\par
The first exact solution for the metric~\eqref{Bianchi-I} in empty space was found by Kasner~\cite{Kasner},
in fact before Friedmann's works. The Kasner solution has been rediscovered many times, and its importance was only appreciated later. 
A particular form of Kasner's solution for the Bianchi-I universe had been discovered earlier in Refs.~\cite{Weyl,Levi-Civita}.  
In 1963, Khalatnikov and Lifshitz began employing the Bianchi universes (Bianchi-I, in particular) for studying the problem of the cosmological singularity~\cite{Lif-Khal}.
At the end of the 1960s, Belinski, Khalatnikov, and Lifshitz discovered the phenomenon of the oscillatory approach to the cosmological singularity~\cite{BKL,BKL1}.
Using the Hamiltonian formalism, Misner referred to this phenomenon as the Mixmaster
Universe~\cite{Misner}.
Later, it was understood that the dynamics of the universe becomes chaotic when the singularity is approached~\cite{chaos,chaos1}.
Finally, the connection between the chaotic behavior in cosmological models and
infinite-dimensional Lie algebras~\cite{Kac} was discovered at the beginning of the
new millennium~\cite{DHN,DHN1,DHN2}. Thus, the study of Bianchi cosmologies, starting from the empty Bianchi-I model,
has already led to some unexpected discoveries.
In the meantime, exact solutions for the Bianchi-I universe filled with matter
were also studied.
First of all, we mention the Heckmann--Schucking solution, which describes the Bianchi-I universe filled with dust~\cite{Heck-Schuck}. 
The Heckmann--Schucking universe behaves as the Kasner universe at the beginning of the evolution,
after which isotropization takes place so that the universe approaches an isotropic Friedmann
regime while expanding.
The generalization of the Heckmann--Schucking solution to other kinds of isotropic perfect fluids
was obtained by Jacobs~\cite{Jacobs0,Jacobs,Jacobs1}, and more generalizations were found
recently~\cite{Khal-Kam,Kam-Ming}. 
\par
While exact solutions with a highly anisotropic geometry do exist even in empty space or in a universe
filled with isotropic matter, the situation becomes even more riveting in the presence of spatially homogenous but anisotropic matter.
A magnetic field is an interesting example of such a source.
Indeed, papers devoted to the Bianchi-I universes with a spatially homogeneous magnetic field oriented along one of the coordinate
axes in Eq.~\eqref{Bianchi-I} were published already in the early 60's, sometimes with additional types of matter~\cite{Rosen,Rosen1,Dorosh,Shikin,Thorne,Jacobs}.  
\par
The interest in solutions involving magnetic fields is not purely academic.
The existence of large-scale magnetic fields in our universe is an important and
enigmatic phenomenon (see, {\em e.g.},~Ref.~\cite{Barrow-magn}).
Their origin is not known and is being widely discussed.
Thus, even if it is unrealistic to describe the present-day universe with the Bianchi-I metric, models where this metric is sustained
by a magnetic field could shed some light on processes that occurred in the very early universe.
In spite of the long history of studies of cosmic magnetic fields in general and in the
Bianchi-I universe in particular, to our knowledge, there is no detailed description of the qualitative behavior of the corresponding solutions.
Besides, it would be interesting to try and bridge the established (and sometimes implicit) knowledge about solutions with magnetic fields with innovative approaches to the
problem of the singularities in gravity and theoretical cosmology.
In this paper, we shall try to fill some gaps and look for new ways in this direction.
\par
The structure of the paper is the following: in Sec.~\ref{s:BianchiIB}, we consider the Bianchi-I model
with a spatially homogeneous magnetic field directed along the $z$ axis of Eq.~\eqref{Bianchi-I}. 
At the beginning and end of the cosmological evolution, the universe is in Kasner regimes.
We establish the relationship between the parameters of these two regimes and show
that it coincides with the one in the empty Bianchi-II universe. 
We then consider two other models in Sec.~\ref{s:BianchiIBF}, one in which dust is added
and one with an additional massless scalar field (or, in other words, stiff matter).
Sec.~\ref{s:sing} is devoted to the possibility of crossing the singularity in these models,
and Sec.~\ref{s:conc} contains some concluding remarks. 
\section{Bianchi-I model with spatially homogeneous magnetic field}
\label{s:BianchiIB}
Let us consider the Bianchi-I universe with the metric in Eq.~\eqref{Bianchi-I}.
The Lagrangian of the electromagnetic field is (in the Gaussian system)~\cite{Land-Lif}
\begin{equation}\label{elec}
L_{\rm em} = -\frac{1}{16 \pi}\,F_{ik}\,F^{ik}
\ ,
\end{equation}  
where Latin indices are four-dimensional, $i=0,\ldots,3$, and $x^i=(t,x,y,z)$
in Eq.~\eqref{Bianchi-I}. 
The energy-momentum tensor of the electromagnetic field with the Lagrangian~\eqref{elec} has the form
\begin{equation}\label{elec1}
T_{\ k}^{i}
= \frac{1}{4\,\pi} \left(-F^{il}\,F_{kl}
+ \frac{1}{4} \, \delta_k^i\, F_{lm}\, F^{lm}\right).
\end{equation}
In particular, we consider the case with no electric field and a homogeneous magnetic field
along $z$.
Hence, the only non-zero component of the electromagnetic field tensor is $F_{12}$. 
The sourceless Maxwell equation for the electromagnetic field reads
\begin{equation}\label{elec2}
F_{[ij;k]} = 0,
\end{equation} 
where semicolons denote covariant derivatives in the metric~\eqref{Bianchi-I} and square brackets imply antisymmetrization.
Since there is no torsion, the symmetric connection terms cancel out and
Eq.~\eqref{elec2} simplifies to $F_{[ij,k]} = 0$.
Choosing the triplet of indices $0$, $1$, and $2$, we see that $F_{12,0} = 0$,
which means that $F_{12}$ is constant. 
For the diagonal spacetime~\eqref{Bianchi-I}, the only non-vanishing component of the fully contravariant electromagnetic field tensor is thus given~by 
\begin{equation}\label{elec5}
F^{12} = g^{11}\,g^{22}\,F_{12} \propto a^{-2}\,b^{-2}.
\end{equation}
That means that all the contributions to the mixed components of the energy-momentum tensor
in Eq.~\eqref{elec1} are proportional to $a^{-2}\,b^{-2}$.
On choosing a convenient parametrization, we can then write 
\begin{equation}\label{elec6}
T_{\ 0}^0 = -T_{\ 1}^1 = -T_{\ 2}^2 = T^3_{\ 3} = \frac{B_0^2}{a^2 b^2},
\end{equation}
where $B_0^2$ is a positive constant characterizing the intensity of the magnetic field.
It is easy to see that the trace $T$ of the energy-momentum tensor~\eqref{elec6} vanishes,
as it should. 
\par
It is convenient to rewrite the Einstein equations, 
\begin{equation}\label{Ein}
G^i_{\ j} \equiv R^i_{\ j} - \frac{1}{2}\,{\cal R}\,\delta^i_{\ j} = T^i_{\ j},
\end{equation}
by parameterizing the scale factors as
\begin{equation}
\begin{gathered}
\label{magn9}
a(t) = R(t) \, \e^{\alpha(t)+\beta(t)}\ ,
\\
b(t) = R(t) \, \e^{\alpha(t)-\beta(t)} ,
\\
c(t) = R(t) \, \e^{-2\alpha(t)} 
\ .
\end{gathered} \end{equation}
The components of the Ricci tensor then read
\begin{eqnarray}
 R_{\ 0}^0
&\!\!=\!\!&
- 3 \frac{\ddot{R}}{R} -6 \dot{\alpha}^2 - 2 \dot{\beta}^2,
 \label{magn10}
 \\
 R_{\ 1}^1
 &\!\!=\!\!&
- \frac{\ddot{R}}{R} - 2 \frac{\dot{R}^2}{R^2} - \ddot{\alpha} - 3 \dot{\alpha} \frac{\dot{R}}{R} - \ddot{\beta} - 3 \dot{\beta} \frac{\dot{R}}{R},
 \label{magn11}
 \\
 R_{\ 2}^2
 &\!\!=\!\!&
 - \frac{\ddot{R}}{R} - 2 \frac{\dot{R}^2}{R^2} - \ddot{\alpha} - 3 \dot{\alpha} \frac{\dot{R}}{R} + \ddot{\beta} + 3 \dot{\beta} \frac{\dot{R}}{R},
 \label{magn12}
 \\
R_{\ 3}^3
&\!\!=\!\!&
- \frac{\ddot{R}}{R} - 2 \frac{\dot{R}^2}{R^2} + 2 \ddot{\alpha} + 6 \dot{\alpha}\frac{\dot{R}}{R},
\label{magn13}
\end{eqnarray}
and Einstein's equations are given by
\begin{equation}\label{magn14}
G_{\ 0}^0 = -G^1_{\ 1} = -G^2_{\ 2} = G_{\ 3}^3
= \frac{B_0^2}{R^4}\, \e^{-4\alpha(t)}.
\end{equation}
Note also that the scalar curvature is
\begin{equation}\label{Ricci}
{\cal R}
\equiv
R^i_{\ i}
=
- 6 \frac{\ddot{R}}{R} - 6 \frac{\dot{R}^2}{R^2} - 6\dot{\alpha}^2 - 2 \dot{\beta}^2,
\end{equation}
which must vanish since $T=0$.
\par
Taking the difference of the mixed ${^1}_1$ and ${^2}_2${--}components of Einstein's equations~\eqref{magn14}
with the expressions~\eqref{magn11} and~\eqref{magn12} for the Ricci tensor, we obtain
\begin{equation}\label{magn19}
\dot{\beta} = \frac{\beta_0}{R^3},
\end{equation}
which is just like in Kasner's and Heckmann-Schucking's solutions~Ref.~\cite{Khal-Kam}.
Likewise, combining Eqs.~\eqref{magn11}, \eqref{magn12}, and~\eqref{magn13}, {\em viz.},
$R_{\ 1}^1+R_{\ 2}^2-2R_{\ 3}^3$, yields
\begin{equation}
\ddot{\alpha}+3\dot{\alpha}\frac{\dot{R}}{R} = \frac{2\,B_0^2}{3\,R^4}\, \e^{-4\alpha(t)}.
\label{magn22}
\end{equation}
Moreover, the combination $R_{\ 1}^1+R_{\ 2}^2+2 R_{\ 3}^3$ provides
\begin{equation}
\ddot{\alpha}+3\dot{\alpha}\frac{\dot{R}}{R} = 2\frac{\ddot{R}}{R}+4\frac{\dot{R}^2}{R^2}.
\label{magn24}
\end{equation}
By multiplying for the spatial volume, $V(t) \equiv R^3(t)$, the equation above can be
rewritten in a more convenient form and then integrated, resulting in
\begin{equation}\label{magn28}
\dot{\alpha} = 2 \frac{\dot{R}}{R}+\frac{\alpha_0}{R^3},
\end{equation}
where $\alpha_0$ is a constant.
Combining Eqs.~\eqref{magn22} and \eqref{magn24} leads~to
\begin{equation}\label{magn29}
\frac{1}{R^3}\,\frac{d^2R^3}{dt^2}
= \frac{B_0^2}{R^4}\, \e^{-4\alpha(t)}.
\end{equation}
Thus, we can determine $\alpha(t)$ for a given $R(t)$, and we further notice that the second time
derivative of the spatial volume must always be positive.
Substituting Eqs.~\eqref{magn10}, \eqref{Ricci}, \eqref{magn19}, \eqref{magn28}, and~\eqref{magn29}
into the ${^0}_0${--}component of the Einstein Eqs.~\eqref{magn14}, after some manipulation,
we obtain an equation for the spatial volume which reads
\begin{equation}\label{magn33}
V\,\ddot{V}+\dot{V}^2+4 \alpha_0 \dot{V} + 3\alpha_0^2 + \beta_0^2 = 0.
\end{equation}
Remarkably, this equation is integrable, but it is instructive to perform a qualitative analysis first. 
\par
Let us point out that not all solutions of Eq.~\eqref{magn33} solve
the complete system of Einstein's and Maxwell's equations.
Since $V(t)$ should always be nonnegative and $\ddot{V}>0$ from Eq.~\eqref{magn29},
we must have $\alpha_0\, \dot V<0$ with $\alpha_0\neq 0$, and
\begin{equation}\label{magn41}
-2\alpha_0 -\sqrt{\alpha_0^2-\beta_0^2} \, 
\leq \, \dot{V} \leq -2\alpha_0+\sqrt{\alpha_0^2-\beta_0^2},
\end{equation}
which is only possible for $\alpha_0^2 \geq \beta_0^2$.
\subsection{Contracting universe}
Let us first consider $\alpha_0>0$, corresponding to a contracting universe
with $\dot V<0$. One can start at a certain moment in time with a positive value of $V$ and a negative value of
$\dot{V}$ satisfying the inequality $\alpha_0^2 \geq \beta_0^2$. Since $\ddot{V}>0$, the time derivative of $V(t)$ grows, remaining negative, and the absolute value of $\dot{V}$ decreases, always satisfying the constraint
\begin{equation}\label{magn43}
\bigl| \dot{V} \bigr| \, \geq \, 2 \alpha_0 - \sqrt{\alpha_0^2-\beta_0^2}
\equiv W_1.
\end{equation}
The universe will therefore reach the singularity characterised by $V = 0$ in
a finite period of time.
\par
Now, we can consider two times $t_1$ and $t_2$ such that $V(t_1) = 0$ and $\dot{V}(t_2) = -W_1$,
and we would like to understand which occurs first. 
Suppose that $t_1 < t_2$, so that $V$ vanishes while $\dot{V}$ still satisfies the inequality~\eqref{magn41}
with $|\dot{V}|$ larger than the critical value $W_1$ from Eq.~\eqref{magn43}.
The time $t_1$ cannot be infinite because the absolute value of the time derivative is larger than $W_1$, and the spatial volume
$V(t)$ reaches zero in a finite period of time. In particular, one can approximate the volume function for $t\lesssim t_1$ with the expression 
\begin{equation}\label{dif3}
V \simeq \gamma \,(t_1-t)^{\lambda},
\end{equation}  
where $\gamma$ and $\lambda$ are positive constants.
For $\lambda > 1$, the velocity becomes 
\begin{equation}\label{dif4}
\dot{V} \simeq -\lambda\, \gamma \,(t_1-t)^{\lambda -1} \to 0 \quad {\rm for} \ t \to t_1,
\end{equation}
which contradicts the condition~\eqref{magn43}. 
On the other hand, if $\lambda < 1$, the velocity in Eq.~\eqref{dif4} diverges for $t\to t_1$, which
violates the bound~\eqref{magn41}.
The only choice left is $\lambda=1$ and we need to include another term in the expansion around $t_1$,
namely
\begin{equation} \label{dif5}
V \simeq \gamma \,(t_1-t) + \eta_1 \,(t_1 -t)^{\mu}
\ ,
\end{equation}
where $\eta_1$ is a positive constant and $\mu > 1$. 
Substituting the corresponding expressions for $V$, $\dot{V}$, and $\ddot{V}$
into Eq.~\eqref{magn33}, we find
\begin{widetext}
\begin{equation}\label{dif8}
\mu \, \eta_1 \,\left(\mu-1\right) \left[\gamma (t_1-t) + \eta_1 (t_1-t)^{\mu}\right]
(t_1-t)^{\mu-2}
\simeq 
-\left[\gamma +\mu\, \eta_1 \, (t_1-t)^{\mu-1} -2\alpha_0 \right]^2
+\alpha_0^2 - \beta_0^2
\ . 
\end{equation}
\end{widetext}
The leading term in the left-hand side above behaves as $(t_1-t)^{\mu-1}$,
which vanishes for $t\to t_1$.
The leading term in the right-hand side is instead a constant, which should therefore vanish, to wit
\begin{equation}\label{dif9}
\gamma^2-4\alpha_0\gamma +3\alpha_0^2 + \beta_0^2 = 0.
\end{equation}
One of the solutions of this equation is $\gamma = W_1$, 
which means that $\dot{V}$ reaches the critical value $W_1$ at the same time when the 
volume $V(t)$ vanishes, so that $t_1=t_2$. 
Next, we equate terms of order $(t_1-t)^{\mu-1}$ in Eq.~\eqref{dif8},
which gives 
\begin{equation}\label{dif12}
\mu = 1 + \frac{2\sqrt{\alpha_0^2-\beta_0^2 \,}}{2\alpha_0-\sqrt{\alpha_0^2-\beta_0^2 \,}}.
\end{equation}
One can see that $1 < \mu \leq 3$, 
whereas the constant $\eta_1$ takes different values depending on the initial conditions.
\par 
We could also consider the case of $\dot V$ reaching the critical value $-W_1$
at the moment $t_2<t_1 $ while the volume $V(t_2) > 0$.
A simple analysis similar to the above shows that this case is excluded.
Thus, we can say that the universe hits the singularity 
$V(t)=0$ at some finite time $t=t_1$ when the velocity $\dot{V}$ reaches the critical value
$-W_1$ for any contracting evolution.  
\par 
From the (approximate) evolution law of the volume, we can determine the anisotropy factors
$\alpha(t)$ and $\beta(t)$.
Using Eq.~\eqref{magn28}, we get
\begin{equation}\label{dif15}
\alpha(t) \simeq
\left(\frac23 -\frac{\alpha_0}{2\alpha_0-\sqrt{\alpha_0^2-\beta_0^2}}\right)\ln(t_1-t). 
\end{equation} 
Analogously, from Eq.~\eqref{magn19}, we immediately find 
\begin{equation}\label{dif14}
\beta(t) = \beta_0 \int \frac{d t}{V} \simeq
-\frac{\beta_0\,\ln(t_1-t)}{2\alpha_0-\sqrt{\alpha_0^2-\beta_0^2}}.
\end{equation}
For definiteness, let us set $\beta_0 \geq 0$.
Using the definitions~\eqref{magn9}, we can write the three scale factors 
in the Kasner form
\begin{equation}\begin{gathered}\label{dif19}
a(t) \sim
\left(t_1-t\right)^{p_1}, 
\\
b(t) \sim
\left(t_1-t\right)^{p_2}, \\
c(t) \sim
\left(t_1-t\right)^{p_3},
\end{gathered} \end{equation}
where \begin{eqnarray}\label{dif20}
p_1
&\!\!=\!\!& 
\frac{\alpha_0-\beta_0-\sqrt{\alpha_0^2-\beta_0^2}}{2\alpha_0-\sqrt{\alpha_0^2-\beta_0^2}} < 0,
\nonumber
\\
p_2
&\!\!=\!\!&
\frac{\alpha_0+\beta_0-\sqrt{\alpha_0^2-\beta_0^2}}{2\alpha_0-\sqrt{\alpha_0^2-\beta_0^2}} > 0,
\\
p_3
&\!\!=\!\!&
\frac{\sqrt{\alpha_0^2-\beta_0^2}}{2\alpha_0-\sqrt{\alpha_0^2-\beta_0^2}} >0.
\nonumber
\end{eqnarray}
It is straightforward to check that the exponents $p_1$, $p_2$, and $p_3$ indeed satisfy the Kasner relations,
{\em i.e.},
\begin{equation}
p_1+p_2+p_3 = p_1^2+p_2^2+p_3^2=1,
\label{dif21}
\end{equation}
which means that the presence of the magnetic field does not change the character of the singularity.
The reason for such a behavior is not hard to guess.
Substituting the expressions for $V$ and $\ddot{V}$ into Eq.~\eqref{magn29},
one obtains that the magnetic field contributes to Einstein's equations as
\begin{equation}\label{dif22}
\frac{B_0^2}{a^2 b^2}
\sim \frac{\mu\,\eta_1}{\gamma} \, (\mu-1)\,(t_1-t)^{\mu-3}
\ ,
\end{equation}
where $\mu-3 > -2$.
Therefore, the term~\eqref{dif22} is weaker than the anisotropy, which contributes a term of order
$(t_1-t)^{-2}$ and dominates near the singularity. The presence of matter less stiff than stiff matter is well known to
leave the Kasner type of singularity unaffected~\cite{Bel-Khal}.
\subsection{Expanding universe}
Let us consider now the expanding universe.
In this case, the constant $\alpha_0<0$ and the time derivative $\dot{V}$ will correspondingly be positive.
The expansion will last for $t\to\infty$ with the time derivative $\dot{V}$ approaching the critical value 
\begin{equation}\label{V2}
W_2 \equiv -2 \alpha_0 + \sqrt{\alpha_0^2-\beta_0^2} > 0.
\end{equation}
Thus, the behavior of $V(t)$ can be approximated in the limit $t\to\infty$ as 
\begin{equation}\label{dif23}
V \simeq W_2\,t-\eta_2\,t^{\nu} ,
\end{equation} 
where $\eta_2$ is a positive constant and $0 < \nu < 1$.
On substituting the corresponding expressions for $V$, $\dot{V}$, and $\ddot{V}$ in Eq.~\eqref{magn33}, we find again
\begin{equation}\label{dif26}
\nu = 1 + \frac{2\sqrt{\alpha_0^2-\beta_0^2\, }}{2 \alpha_0 - \sqrt{\alpha_0^2-\beta_0^2}} ,
\end{equation}
with $\dfrac13 \leq \nu < 1$.
The anisotropy factors then read
\begin{align}\label{dif29}
\alpha(t) &= \left(\frac23+\frac{\alpha_0}{\sqrt{\alpha_0^2-\beta_0^2}-2\alpha_0} \right) \ln{t}, \\
\label{dif28}
\beta(t) &= \frac{\beta_0\, \ln{t}}{\sqrt{\alpha_0^2-\beta_0^2}-2\alpha_0}.
\end{align}
Correspondingly, the scale factors take again the Kasner form
\begin{equation}\label{dif30}
a(t) \sim 
t^{p_1}, \quad 
b(t) \sim 
t^{p_2}, \quad 
c(t) \sim 
t^{p_3},
\end{equation}
where 
\begin{eqnarray}
p_1 &\!\!=\!\!&
\frac{\sqrt{\alpha_0^2-\beta_0^2}-\alpha_0+\beta_0}{\sqrt{\alpha_0^2-\beta_0^2}-2\alpha_0}, 
\nonumber \\
p_2 &\!\!=\!\!&
\frac{\sqrt{\alpha_0^2-\beta_0^2}-\alpha_0-\beta_0}{\sqrt{\alpha_0^2-\beta_0^2}-2\alpha_0}, 
\label{dif31} \\
p_3 &\!\!=\!\!&
-\frac{\sqrt{\alpha_0^2-\beta_0^2}}{\sqrt{\alpha_0^2-\beta_0^2}-2\alpha_0},
\nonumber
\end{eqnarray} 
which also satisfy the Kasner relations~\eqref{dif21}.
The presence of the magnetic field does not influence the asymptotic structure of the metric at $t \to\infty$,
which therefore does not isotropize, unlike the Heckmann-Schucking solution with dust~\cite{Heck-Schuck,Khal-Kam}.
The reason for this phenomenon is clear. 
The energy density of the magnetic field at $t \to\infty$ is given by
\begin{equation}\label{dif32}
\frac{B_0^2}{a^2b^2} \simeq \nu\,(\nu-1)\,\frac{\eta_2}{W_2}\,t^{\nu-3},  
\end{equation} 
where $\nu-3 < -2$.
Hence it remains weaker than the anisotropy term. 
\par
Now we can try to understand what happens with the expanding universe in its distant past.
One can suppose that it was born from the initial singularity at $t=0$, when its volume $V(0)=0$,
and its time derivative had the smallest critical value
\begin{equation}
\dot V(0) = W_3 \equiv -2\alpha_0-\sqrt{\alpha_0^2-\beta_0^2}>0.
\label{dif33}
\end{equation}
In this case, the scale factors will be of the form in Eq.~\eqref{dif30} with the Kasner exponents
\begin{eqnarray} \label{dif34}
p'_1 
&\!\!=\!\!&
\frac{\alpha_0-\beta_0+\sqrt{\alpha_0^2-\beta_0^2}}{2\alpha_0+\sqrt{\alpha_0^2-\beta_0^2}}, \nonumber \\
p'_2
&\!\!=\!\!&
\frac{\alpha_0+\beta_0+\sqrt{\alpha_0^2-\beta_0^2}}{2\alpha_0+\sqrt{\alpha_0^2-\beta_0^2}}, \\
p'_3 
&\!\!=\!\!&
- \frac{\sqrt{\alpha_0^2-\beta_0^2}}{2\alpha_0+\sqrt{\alpha_0^2-\beta_0^2}}.
\nonumber
\end{eqnarray} 
They again satisfy the Kasner relations~\eqref{dif21}. To establish a relation between the set of Kasner indices at the beginning
and at the end of the evolution, it is convenient to use the
Lifshitz--Khalatnikov parametrization~\cite{Lif-Khal}. 
If the Kasner indices are ordered as 
\begin{equation}
p_1 \leq p_2 \leq p_3,
\label{dif35}
\end{equation}
they can be represented by means of a real parameter $u\geq 1$ according to
\begin{equation} \begin{gathered}
p_1 =
-\frac{u}{1+u+u^2},
\\
p_2 =
\frac{1+u}{1+u+u^2},
\label{dif36} \\
p_3 =
\frac{u\,(1+u)}{1+u+u^2}.
\end{gathered} \end{equation}
The ordering~\eqref{dif35} can be obtained, for example, by 
setting the anisotropy parameters
\begin{equation}
\alpha_0 < 0,
\quad
\beta_0 < 0,
\quad
|\beta_0| < \frac35\,|\alpha_0|.
\label{dif37}
\end{equation}
In particular, this choice implies that the universe expands in the $y$ and $z$ directions,
but does so more rapidly along the direction $z$ of the magnetic field, while it contracts along the $x$ axis.
Combining Eqs.~\eqref{dif34} and~\eqref{dif37}, we obtain
\begin{equation}
u' = \frac{p_3'}{p_2'} =
\frac{\sqrt{\alpha_0^2-\beta_0^2}}{|\alpha_0|+|\beta_0|-\sqrt{\alpha_0^2-\beta_0^2}}.
\label{dif38}
\end{equation}
It is convenient to introduce the parameter 
$\xi = \dfrac{|\beta_0|}{|\alpha_0|}$,
which, when plugged into Eq.~\eqref{dif38}, results in the relation
\begin{equation}
\xi = \frac{(u'+1)^2-u'^2}{(u'+1)^2+u'^2}.
\label{dif39}
\end{equation} 
Note that, if $\xi$ satisfies the conditions~\eqref{dif37}, then $1 < u' < \infty$. 
Let us now look at the Kasner exponents~\eqref{dif31} describing the final stage of cosmological evolution. 
In this case the order of the exponents is 
\begin{equation}
p_3 \leq p_1 \leq p_2.
\label{dif40}
\end{equation}
Correspondingly, we can represent them as 
\begin{equation} \begin{gathered}
p_1 =
\frac{1+u}{1+u+u^2}, \quad \label{dif41}
\\
p_2 =
\frac{u\,(1+u)}{1+u+u^2}, \\
p_3 =
-\frac{u}{1+u+u^2}.
\end{gathered} \end{equation}
Thus,
\begin{equation}
u = \frac{p_2}{p_1} =
\frac{\sqrt{\alpha_0^2-\beta_0^2}-\alpha_0-\beta_0}{\sqrt{\alpha_0^2-\beta_0^2}-\alpha_0+\beta_0}.
\label{dif42}
\end{equation}
Substituting the formulas~\eqref{dif38} and \eqref{dif39} into the equation above, we find
\begin{equation}
u = \frac{1+u'}{u'} < 2.
\label{dif43}
\end{equation}
Inversely,
\begin{equation}
u' = \frac{1}{u-1}.
\label{dif44}
\end{equation}
\par
For the inverse evolution towards the singularity, as one usually analyzes the oscillating approach towards
the cosmological singularity~\cite{BKL,Bel-Hen}, one can see that the universe passes through two transformations
in the transition from the phase described by the parameter $u$ to the phase described by $u'$,
according to Eq.~\eqref{dif44}. The first transformation is characterized by the shift $u \to u-1$,
in which the roles of the $x$ and $z$ axes, corresponding to the exponents $p_1$ and $p_3$, are exchanged.
This transformation is called ``change of the Kasner epoch''~\cite{BKL}.
As a result of this transformation, we arrive at a value of the parameter $u-1<1$, see Eq.~\eqref{dif43}.
The next transformation is defined by 
\begin{equation}
u-1 \to \dfrac{1}{u-1\, },
\end{equation}
which exchanges the roles of the axes $y$ and $z$, and is called  ``change of Kasner era''. 
\par
We see that our solution displays a transition between two Kasner regimes
chacterized by the same law found in an empty Bianchi-II universe.
For a detailed description of the dynamics in the Bianchi-II universe in General Relativity
and other gravity models see Ref.~\cite{Giani-Kam}. In Bianchi-VIII or Bianchi-IX models, the universe passes through an infinite series
of changes of the Kasner epochs and eras.
In our case, for the choice of parameters in Eq.~\eqref{dif37}, the law of the transformation for the Kasner exponents~\eqref{dif44} includes one change of Kasner epoch and one change of Kasner era. 
\par
One can also consider the opposite relation between the anisotropy parameters, to wit
\begin{equation}
\alpha_0 < 0, \quad \beta_0 < 0,
\quad \frac35\,|\alpha_0| < |\beta_0| < 1,
\label{dif45}
\end{equation}
or, in other words, 
$3/5 < \xi < 1$, 
and we have
$u' = p_2' / p_3'$,
which corresponds~to
\begin{equation}
\xi = \frac{(u'+1)^2-1}{(u'+1)^2+1},
\label{dif50}
\end{equation}
and $u = u'+1$
or, inversely, $u'=u-1$.
The last two relations show that we now have only a change of Kasner epoch.
\subsection{Exact evolution}
As we already pointed out, in addition to the qualitative analysis given in the previous subsections, Eq.~\eqref{magn33} in principle is integrable.
We can cast it in the form
\begin{equation}\label{exact}
\frac{d}{dt}\left(V\,\dot{V}+4\alpha_0\, V\right) = -3\alpha_0^2-\beta_0^2,
\end{equation}
which can be integrated
by defining $V = X t$, leading to 
\begin{widetext}
\begin{equation}\label{magn48}
\frac12\ln\left(1+\frac{X^2+4\alpha_0X}{3\alpha_0^2+\beta_0^2}\right)
+\frac{\alpha_0}{\sqrt{\alpha_0^2-\beta_0^2}}
\left[\ln\biggl(1+\frac{X}{2\alpha_0+\sqrt{\alpha_0^2-\beta_0^2}}\biggr)
- \ln\biggl(1+\frac{X}{2\alpha_0-\sqrt{\alpha_0^2-\beta_0^2}}\biggr)
\right]
=
\ln\left(\frac{t_0}{t}\right). 
\end{equation}
\end{widetext}
where integration constants are chosen so that $t_0>0$ is the time at which the volume $V(t)$, hence $X(t)$, vanishes.
\par
Unfortunately, Eq.~\eqref{magn48} cannot be inverted to find $X(t)$ as a function of $t$.
Thus, we cannot use this equation to find exact expressions for the anisotropy factors.
Notwithstanding, one can use it to study the approach to the singularity for $t\to t_0$ when $X(t) \to 0$.
Expanding the left-hand side of Eq.~\eqref{magn48}, we see that the leading term is proportional to $X^2$, and the equation reduces to 
\begin{equation}\label{magn49}
X^2 \simeq 2\left(3\alpha_0^2 + \beta_0^2 \right)\frac{t_0-t}{t_0}.
\end{equation} 
This implies
\begin{equation}
\label{magn50}
V \simeq \sqrt{2\bigl(3\alpha_0^2 + \beta_0^2 \bigr)t_0\,(t_0-t)} 
\end{equation} and $\dot{V}  \to -\infty$, 
which obviously breaks the inequality~\eqref{magn41}.
The approximate solution~\eqref{magn49} therefore corresponds to initial conditions which
are incompatible with the the existence of a homogenous magnetic field. 
\par
However, the exact solution~\eqref{magn48} admits a different regime corresponding
to a volume singularity given by $t \to 0$ with $X(0)$ finite.
To describe this regime, one can confront the terms proportional to $\ln t$ in the
left-hand side and the right-hand side of Eq.~\eqref{magn48}.
In this case, the function $X(t)$ tends to a positive constant which can be found
from a quadratic equation and corresponds to the extremal points of the
inequality~\eqref{magn41}. 
Thus, the regime of the approach to the singularity described above is compatible
with the general solution of Eq.~\eqref{exact} and arises in the vicinity of the point
$t = 0$. 
\subsection{Anisotropy and relations with the Bianchi-II universe dynamics}
The Kasner solution for the Bianchi-I universe is the regime of maximal anisotropy.
The degree of this anisotropy does not depend on the particular values of the Kasner indices or,
equivalently, on the value of the Lifshitz-Khalatnikov parameter.
Indeed, to explain this fact, let us introduce the dispersion of the set of Kasner's indices
\begin{equation}
\sigma^2 = \left(p_1-\frac13\right)^2+\left(p_2-\frac13\right)^2+\left(p_3-\frac13\right)^2.
\label{disp}
\end{equation}
Using the Kasner relations \eqref{dif21}, we see that $\sigma^2 = 2/3$ and does not depend
on the particular choice of the $p_i$'s.
Thus, at the beginning and end of the evolution described in the preceding subsections,
the universe is maximally anisotropic ({\em i.e.}, anisotropy evolves like $R^{-6}$),
and the presence of the magnetic field is not significant in the very early and in the
very late universe, as we have seen.
However, during  the intermediate stage of evolution, it is the presence of the
magnetic field that drives the transition from one Kasner regime to another.
\par
We have already highlighted that the character of this transition exactly coincides with one
that takes place in an empty Bianchi-II universe.
It is interesting to find the roots of this phenomenon.
The laws of change of Kasner epochs and, sometimes, eras are well known and can
be worked out in two ways.
The first one is to solve exactly the differential equations for the Bianchi-II model
(see, {\em e.g.},~\cite{Giani-Kam,Giani} and references therein). 
These equations acquire a Liouville-like form if logarithmic time is used.
Alternatively, one can study these equations qualitatively, arriving at the same
conclusions concerning the asymptotic regimes~\cite{Land-Lif}. 
Here, we would like to treat the dynamics of the empty Bianchi-II universe
using the same method that was used to study the Bianchi-I universe with magnetic field.
In this way, we shall manage to understand why the laws of transformation between
Kasner regimes coincide in these quite different physical models.
Besides, we shall obtain another simple way to derive the law of transformation for the Bianchi-II universe. 
\par
All of the Bianchi geometries can be represented in a synchronous reference system by a metric of the form 
\begin{equation}
ds^2 = dt^2 -a^2(t)\,\omega^1\otimes\omega^1-b^2(t)\,\omega^2\otimes\omega^2-c^2(t)\,\omega^3\otimes\omega^3,
\label{Bian-g}
\end{equation}   
where $\omega^1$, $\omega^2$, and $\omega^3$ are the basis one-forms dual to the basis of the reciprocal vector fields
which commute with all the Killing vector fields~\cite{Ryan}.
The differences between various Bianchi models are encoded by the structure coefficients
of the Lie algebra of the three Killing vector fields. Moreover, the components of the Ricci tensor for the
metric~\eqref{Bian-g} have the following form:
\begin{equation}
{R^i}_j = {R^i}_j(K) - {P^i}_j,
\label{Ricci-s}
\end{equation}
where ${R^i}_j(K)$ is the part of the Ricci tensor depending on the extrinsic curvature given in Eqs.~\eqref{magn10}{--}\eqref{magn13}. 
The tensor ${P^i}_j$ is constructed instead from the components of the three-dimensional spatial metric
and can be expressed in terms of the scale factors $a(t)$, $b(t)$, and $c(t)$ and of the structure constants
of the Lie algebra of the Killing vector fields.
For the Bianchi-I Lie algebra, the structure constants are zero, hence ${P^i}_j=0$. For the Bianchi-II Lie algebra,
only one structure constant is not zero and can be chosen so that the Lie bracket of the first two Killing vectors
be equal to the third Killing vector.
The non-vanishing components of the tensor ${P^i}_j$ then read
\begin{equation}\label{Ricci-s1}
{P^1}_1 = {P^2}_2= - {P^3}_3 = - \frac{c^2}{2a^2b^2} =-\frac{\e^{-8\alpha(t)}}{2R^2},
\end{equation}
It follows that we can treat an empty Bianchi-II universe as a Bianchi-I universe filled with a particular kind of matter characterised by
the energy-momentum tensor
\begin{equation}
{T^i}_j = {P^i}_j-\frac12 P \delta^i_j \, ,
\label{Ricci-s2}
\end{equation}
with components
\begin{equation}
{T^0}_0 = - {T^1}_1 = - {T^2}_2 = \frac13 \ {T^3}_3 =  \frac{\e^{-8\alpha(t)}}{4R^2}. 
\label{Ricci-s3}
\end{equation}
Substituting these expressions into Einstein's equations for the Bianchi-I universe,
we can study the dynamics of the Bianchi-II universe.
First of all, considering the difference ${R^1}_1-{R^2}_2$, we again obtain the relation \eqref{magn19}.
Then, the combination ${R^1}_1+{R^2}_2- 2{R^3}_3$ yields a relation similar to Eq.~\eqref{magn22}, to wit
\begin{equation}
\ddot{\alpha}+3\dot{\alpha}\frac{\dot{R}}{R}=\frac{\e^{-8\alpha(t)}}{3 R^2}.
\label{Ricci-s4}
\end{equation} 
Combining ${R^1}_1+{R^2}_2 +2{R^3}_3$, we get the relations \eqref{magn24} and \eqref{magn28} again,
and a combination of \eqref{magn24} with \eqref{Ricci-s4} yields
\begin{equation}
\frac{1}{R^3}\frac{d^2 R^3}{dt^2} = \frac{\e^{-8\alpha(t)}}{2R^2},
\label{Ricci-s5}
\end{equation}
which is similar to Eq.~\eqref{magn29}.
To obtain the counterpart of equation~\eqref{magn33}, we use the ${^0}_0${--}component of Einstein equations, and get 
\begin{equation}\label{magn401}
V\ddot{V} + 2 \left(\dot{V}^2 + 4 \alpha_0\,\dot{V}+3 \alpha_0^2 + \beta_0^2 \right) =0.
\end{equation} 
The only difference between Eqs.~\eqref{magn33} and~\eqref{magn401} is the additional factor of $2$
in front of the round bracket of Eq.~\eqref{magn401}.
As follows from Eq.~\eqref{Ricci-s5}, the second time derivative of the volume should be positive,
which implies that the inequality $\alpha_0^2 \geq \beta_0^2$ has to hold. 
\par
\begin{figure}[!th]
\centering
\includegraphics[width=0.49\linewidth]{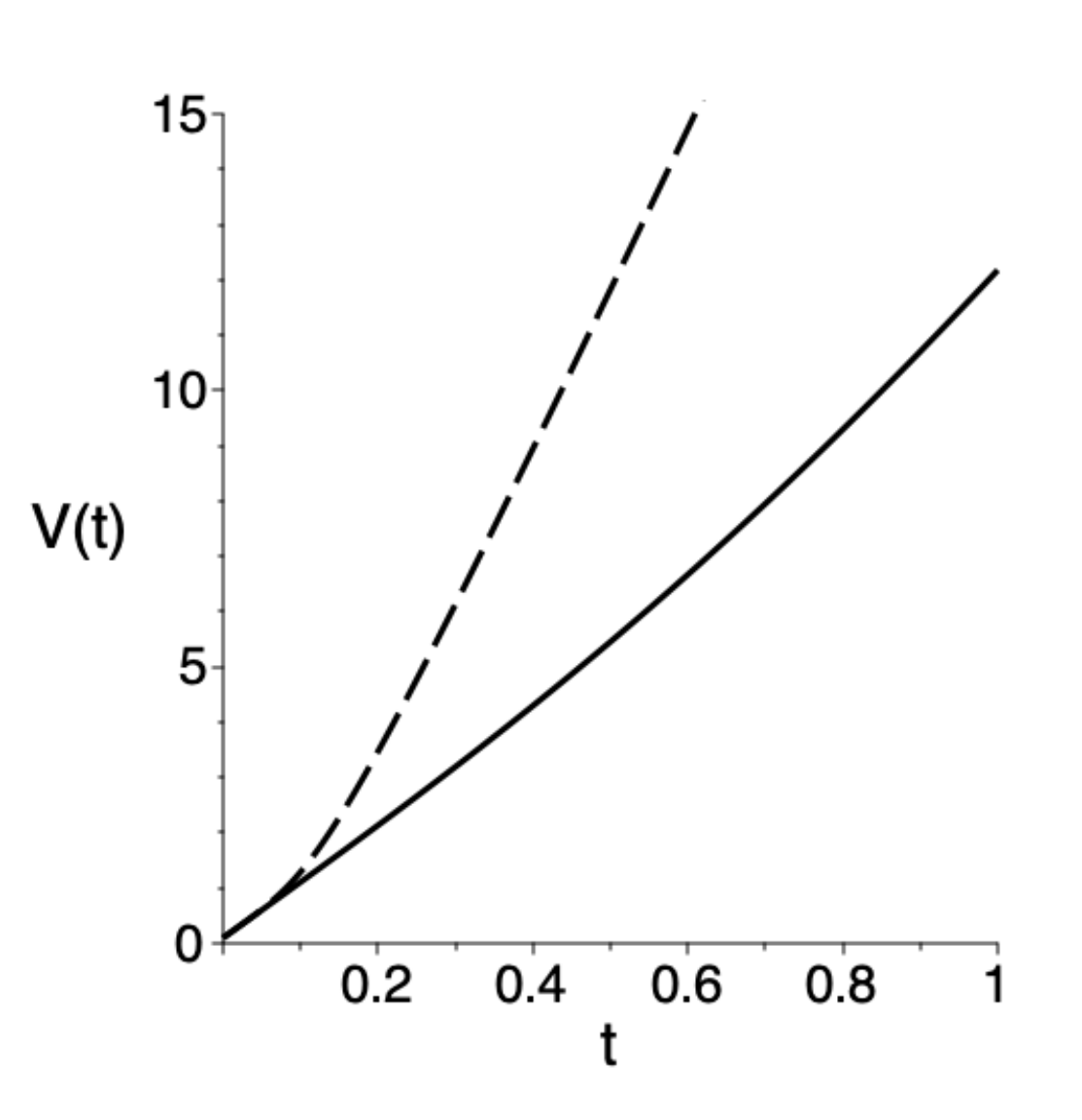}
\includegraphics[width=0.49\linewidth]{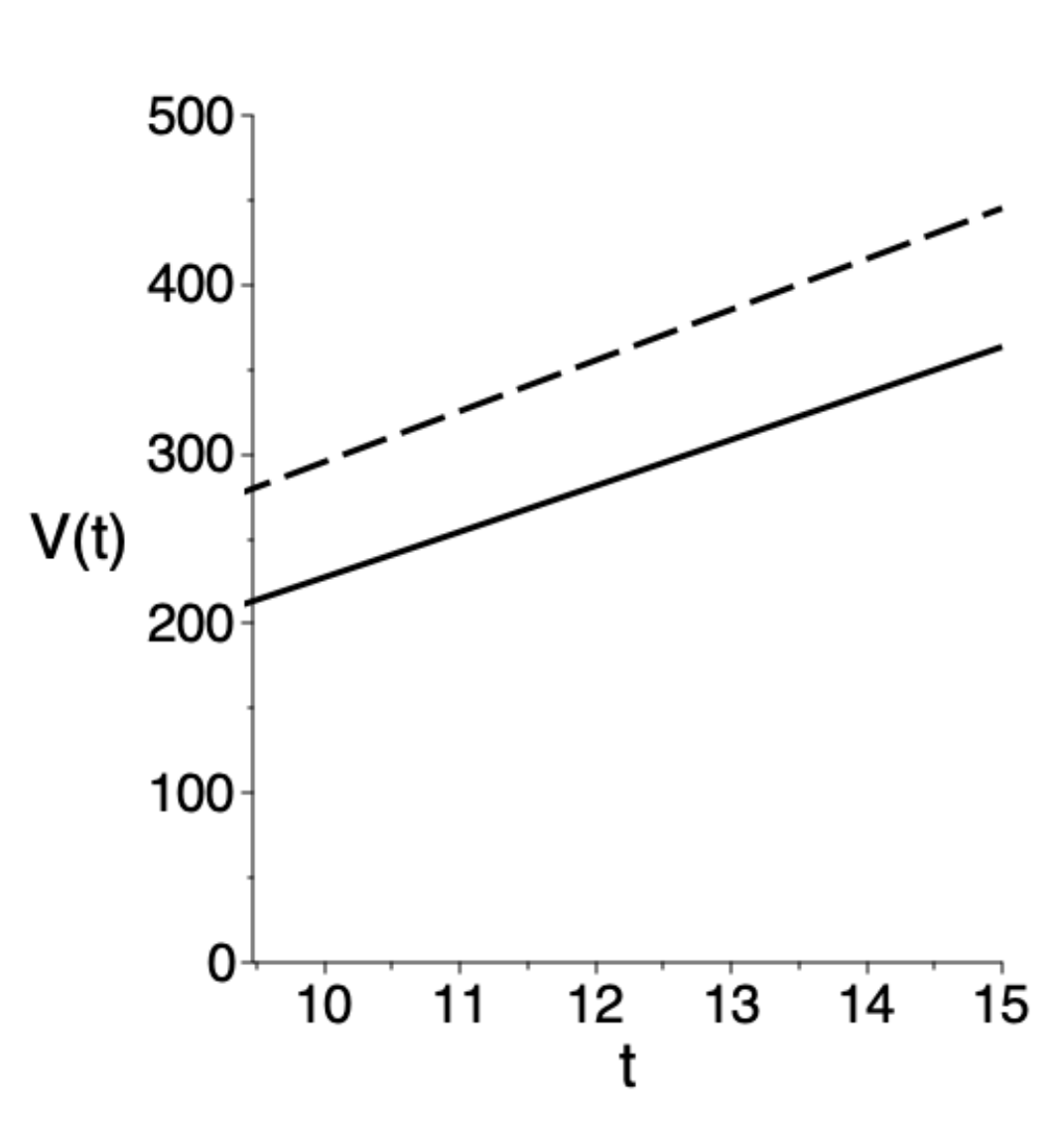}
\caption{Numerical solutions of Eqs.~\eqref{magn33} (solid line) and~\eqref{magn401} (dashed)
for $\alpha_0=-10$, $\beta_0 =1$ and with initial conditions $V(0) = 10^{-6}$, $\dot{V}(0)=10.051 \simeq W_3$
from Eq.~\eqref{dif33}.} 
\label{fig1}
\end{figure}

By repeating all the considerations of the preceding subsections, we can see that the parameters
of the initial and final Kasner regimes coincide with those obtained for the Bianchi-I universe with the magnetic field.
The only difference consists in the fact that the Bianchi-II universe goes out of the initial Kasner regime rapidly
and enters the final Kasner regime early, which is consistent with the Bianchi-I universe filled with a magnetic field.  
Numerical solutions for the associated equations, namely Eqs.~\eqref{magn33} and \eqref{magn401}, are compared in Fig.~\ref{fig1}
for particular values of the parameters and initial conditions.
\section{Bianchi-I models with magnetic field and additional perfect fluids}
\label{s:BianchiIBF}
In the previous section, we analyzed cosmological models just sourced by
a magnetic field. 
Now, we will briefly consider two models in which we add dust and a massless scalar field (stiff matter), respectively.
\subsection{Dust}
The energy density of dust is 
$\rho
= \rho_0 R^{-3}$, 
where $\rho_0$ is a positive constant.
It follows immediately from Einstein's equations that the scalar curvature is then given by
${\cal R} = -\rho$.
It is easy to see that Eq.~\eqref{magn19} does not changed in the presence of dust and the
anisotropy function $\beta$ is still the same. 
Eq.~\eqref{magn24} instead changes because the scalar curvature does not vanish 
and reads
\begin{eqnarray}\label{magn-dust2}
R_{\ 1}^1+R_{\ 2}^2+2 R_{\ 3}^3
&\!\!=\!\!&
-4 \frac{\ddot{R}}{R}-8\frac{\dot{R}^2}{R^2}+6\dot{\alpha} \frac{\dot{R}}{R}+2\ddot{\alpha}
\nonumber \\
&\!\!=\!\!&
2{\cal R} = - \frac{2\rho_0}{R^3}. 
\end{eqnarray}
This equation can be rewritten in the manner of the beginning of this Section, 
which yields, analogously to Eq.~\eqref{magn28},
\begin{equation}\label{magn-dust4}
\dot{\alpha} = 2 \frac{\dot{R}}{R}-\frac{\rho_0 t}{R^3} +\frac{\alpha_0}{R^3}.
\end{equation}
Eq.~\eqref{magn22} is still valid, and we obtain 
\begin{equation}
\frac{B_0^2}{R^4}\, \e^{-4\alpha(t)}
= \frac{1}{R^3}\,\frac{d^2R^3}{dt^2}-\frac{\rho_0}{R^3},
\label{magn-dust5}
\end{equation}
which implies $\dfrac{d^2R^3}{dt^2} > \rho_0$.
By performing the analysis as in the case without dust,
we obtain the equation for the volume  
\begin{equation}\label{magn-dust7} \begin{split}
V\,\ddot{V}
=
&
-\dot{V}^2 + 4\bigl(\rho_0t-\alpha_0\bigr)\dot{V}
-3\alpha_0^2-\beta_0^2
\\
&
+ 3\rho_0
\bigl(2\alpha_0-\rho_0 t\bigr)t,
\end{split} \end{equation} 
which we cannot solve exactly.
\par
An axisymmetric solution is known in the literature~\cite{Dorosh,Shikin,Thorne},
corresponding to $\beta_0 = 0$.
Since this solution is rather cumbersome and implicit,
we shall just undertake a simple qualitative analysis of Eq.~\eqref{magn-dust7}. 
One can see that, when the universe approaches the singularity $V(t)=R^3(t)=0$,
the terms proportional to $\rho_0$ in Eq.~\eqref{magn-dust7} become negligible,
and we come back to the situation without dust.
Thus, we have a Kasner type of singularity with a positive Kasner exponent
in the direction of the magnetic field.
When the volume of the universe grows towards infinity, we encounter the opposite situation.
The term $3\rho_0 t^2$ dominates, since $V(t) \sim t^2$, and we have isotropization
just like in the standard Heckmann-Schucking solution.
\subsection{Massless scalar field}
Let us consider a Bianchi-I universe with magnetic field as in the preceding
sections and a homogeneous massless scalar field, $\phi=\phi(t)$.
For a massless scalar field, the pressure equals the energy density, and the perfect fluid
with such an equation of state is called stiff matter.
The Klein-Gordon equation for $\phi(t)$ in the Bianchi-I universe has the form
\begin{equation}
\ddot{\phi}+3\,\frac{\dot{R}}{R}\,\dot{\phi} = 0,
\label{KG}
\end{equation}
whose general solution is given by
$\dot{\phi} = \dfrac{\tilde{\phi}_0}{R^3}
= \dfrac{\tilde{\phi}_0}{V}$,
where $\tilde{\phi}_0$ is a constant.
One can see that $\dot\phi$ behaves in the same way as the time derivative of the anisotropy
factor $\beta(t)$.
It is convenient to rescale $\tilde{\phi}_0=\sqrt{2}\,\phi_0$, so that the scalar field energy-momentum tensor reads
\begin{equation}
T_{\ 0}^0 = -T_{\ 1}^1 = -T_{\ 2}^2 = -T_{\ 3}^3 = \frac{\phi_0^2}{V^2},
\label{KG2}
\end{equation}
and the scalar curvature ${\cal R} = \dfrac{2\phi_0^2}{V^2}$.
Obviously, the presence of $\phi$ does not change Eq.~\eqref{magn22}.
Moreover, due to the special form of the spatial components in Eq.~\eqref{KG2} and
${\cal R}$, Eq.~\eqref{magn24} also does not change in form.
Therefore, the results in Eqs.~\eqref{magn28} and \eqref{magn29} still hold.
However, the scalar field contributes to the ${^0}_0${--}component of Einstein's equations~\eqref{Ein}.
From Eq.~\eqref{KG2}, we see that $\phi_0$ enters the equation for the volume~\eqref{magn33}
on an equal footing with the parameter $\beta_0$.
Thus, the new equation for $V(t)$ coincides with the formulas in Sec.~\ref{s:BianchiIB} with $\beta_0^2$
replaced by $\beta_0^2+\phi_0^2$.
\par
There is no need to rewrite all of the formulas from Sec.~\ref{s:BianchiIB} explicitly,
but we just focus on the expressions for the scale factors for the expanding universe
in the vicinity of the singularity.
By expanding around $t=0$, we obtain the Kasner form~\eqref{dif30} with
\begin{eqnarray}
p_1
&\!\!=\!\!&
\frac{\alpha_0-\beta_0+\sqrt{\alpha_0^2-\beta_0^2-\phi_0^2}}{2\alpha_0+\sqrt{\alpha_0^2-\beta_0^2-\phi_0^2}}, \nonumber \\
p_2
&\!\!=\!\!&
\frac{\alpha_0+\beta_0+\sqrt{\alpha_0^2-\beta_0^2-\phi_0^2}}{2\alpha_0+\sqrt{\alpha_0^2-\beta_0^2-\phi_0^2}},
\label{KG7} \\ p_3
&\!\!=\!\!&
-\frac{\sqrt{\alpha_0^2-\beta_0^2-\phi_0^2}}{2\alpha_0+\sqrt{\alpha_0^2-\beta_0^2-\phi_0^2}}, \nonumber
\end{eqnarray}
where we have chosen $\alpha_0 < 0$ and $\beta_0 \leq 0$.
It is well known~\cite{Bel-Khal} that the presence of the massless scalar field changes the relations for the Kasner exponents.
While they still satisfy $p_1+p_2+p_3 = 1$,
the sum of their squares is smaller then one.
Namely, using the formulas~\eqref{KG7}, we obtain $p_1^2+p_2^2+p_3^2 = 1-q^2$,
with the parameter 
\begin{equation}
q^2 = \frac{2\phi_0^2}{5\alpha_0^2+4\alpha_0\,\sqrt{\alpha_0^2-\beta_0^2-\phi_0^2}-\beta_0^2-\phi_0^2},
\label{KG10} \end{equation}
satisfying the inequalities
\begin{equation}
0 \leq q^2 < \frac12.
\label{KG11} \end{equation}
It is worth noting that, in the case of a universe filled only with the massless scalar field, the parameter $q^2$
must satisfy the less stringent bounds~\cite{Bel-Khal}
\begin{equation}
0 \leq q^2 \leq \frac23.
\label{KG12}
\end{equation}
One can grasp where the root of the difference lies between Eqs.~\eqref{KG11} and \eqref{KG12}.
The parameter $q^2$ indicates some kind of isotropization for the universe.
The limiting value $q^2 = 2/3$ in Eq.~\eqref{KG12} implies that $p_1=p_2=p_3=1/3$,
{\em i.e.}, that the expansion of the universe is totally isotropic.
The presence of a magnetic field in the $z$ direction makes such a high degree of
isotropization impossible, which is then reflected in Eq.~\eqref{KG11}.
Finally, the Kasner exponents when the volume of the universe tends to infinity are given by
\begin{eqnarray}
p'_1
&\!\!=\!\!&
\frac{\alpha_0-\beta_0-\sqrt{\alpha_0^2-\beta_0^2-\phi_0^2}}{2\alpha_0-\sqrt{\alpha_0^2-\beta_0^2-\phi_0^2}},
\nonumber \\ p'_2
&\!\!=\!\!&
\frac{\alpha_0+\beta_0-\sqrt{\alpha_0^2-\beta_0^2-\phi_0^2}}{2\alpha_0-\sqrt{\alpha_0^2-\beta_0^2-\phi_0^2}},
\label{KG13}
\\ p'_3
&\!\!=\!\!&
\frac{\sqrt{\alpha_0^2-\beta_0^2-\phi_0^2}}{2\alpha_0-\sqrt{\alpha_0^2-\beta_0^2-\phi_0^2}},
\nonumber \end{eqnarray}
so that $(p'_1)^2+(p'_2)^2+(p'_3)^2 = 1-(q')^2$,
where 
\begin{equation}\label{KG15}
(q')^2 = \frac{2\phi_0^2}{5\alpha_0^2-4\alpha_0\sqrt{\alpha_0^2-\beta_0^2-\phi_0^2} -\beta_0^2-\phi_0^2},
\end{equation}
and $(q')^2 < q^2$.
Therefore, in contrast to the isotropization induced by the presence of dust in the Heckmann-Schucking solution,
the value of $q^2$ decreases, going from the singularity towards an infinite expansion, and the degree of
anisotropy increases.
This effect was noticed in the study of the anisotropic universe with the scalar field only ~\cite{Bel-Khal,Khal-Kam}.
\subsection{Some numbers}
While our Bianchi-I universe filled with a spatially homogeneous magnetic field oriented along one of the axes
is substantially a mathematical model and cannot represent a realistic description of our universe,
one can hope that it helps to capture some interesting features of cosmology.
For example, it could describe the evolution of some local patches of the universe.
Thus, it is interesting to try and consider some values for the magnetic field coming from the comparison between
models and observations (see, {\em e.g.}, Refs.~\cite{numbers,numbers0,numbers1,numbers2}), and check
what we can expect in the framework of our solutions.  
\par
According to Ref.~\cite{numbers}, most constraints on cosmological magnetic fields give an upper
bound around $10^{-9}\,$G, assuming coherence on Mpc scales or larger.
Let us suppose that, at the end of the evolution, the magnetic field has such a magnitude and
estimate the value that the magnetic field could have had close to the initial singularity,
or in the very early universe.
One can ask: What does it mean ``close to the singularity''?
For example, we can identify the early time as the beginning of the inflationary expansion.
The time interval to consider will correspondingly be the standard number of $\e$-folds, {\em i.e.},
$N_\e \approx 60$ or, in redshift terms, something like $z \sim 10^{50}$.
In our solution, the magnetic field $B \propto a^{-1}b^{-1}$, so that 
the ratio between the ``early'' and ``late'' values of the magnetic field is 
\begin{equation}
\frac{B_{\rm in}}{B_{\rm fin}} = \frac{a_{\rm fin}}{a_{\rm in}}\, \frac{b_{\rm fin}}{b_{\rm in}}.
\label{ratio1} \end{equation}
If the expansion were isotropic, this ratio would be equal to $z^2 \sim 10^{100}$,
and the initial value of the magnetic field would be huge.
However, anisotropy can change the situation drastically.
Indeed, our evolution represents the transition from one Kasner regime to another.
Consider the situation at the beginning of the evolution, when the scale factor $a(t)$
decreases while $b(t)$ grows.
In this case, 
\begin{equation}
a b \sim t^{p_1+p_2} = t^{1-p_3}
\label{ratio1}
\end{equation} 
and the growth in time of the magnetic field can be very slow if the Kasner index $p_3$
is close to $1$.
Subsequently, the scale factors $a(t)$ and $b(t)$ exchange their roles, though that could
happen (at high values of the parameter $u$) such that the index $p_3$ is still close to $1$.
One can therefore imagine that, during a very large anisotropic expansion, the value of the
magnetic field changes slowly, and the small value of the magnetic field at the end of inflation
is quite compatible with reasonable values in the very early universe.
\section{Singularities and their crossing} \label{s:sing}
The existence of singularities in cosmological models has attracted the attention of
researchers working in General Relativity and its modifications for a long time. 
In fact, the question of the initial singularity was already discussed
in the seminal paper by Robertson~\cite{Robertson}, where the early
development of Friedmann-type cosmologies was reviewed and generalized.
The connection between the presence and sign of the spatial curvature,
the value of the cosmological constant, and the dependence of the pressure
on the scale factor of the universe were studied there in detail with regard
to the appearance of a singularity.
It is interesting that Robertson also considered the scenario of a cyclic evolution,
described by some trigonometric law, in which the universe emerges from a singularity,
expands to the maximum value of its radius, then contracts back, and the process repeats itself indefinitely.  
It appears that a universe which goes through this singularity
did not disturb Robertson too much.
Another type of cosmological singularity, which can arise in the future
for some finite values of the scale factor of the universe and can be rather
soft, was described in Ref.~\cite{Barrow-soft}.
The interest in such singularities essentially increased during the last few
years (see, {\em e.g.}, the review~\cite{my-review} and references therein).
\par
In contrast with the crossing of soft singularities, the idea of crossing the Big~Bang--Big~Crunch
singularity appears rather counter-intuitive.
For many years, the desire to look for models free of such singularities dominated,
although the idea of the possible transition from the Big Crunch to the Big~Bang was studied in some
cosmological models.
First of all, we would like to mention the string or pre-Big Bang scenario~\cite{Gasp-Ven,Lidsey,Gasp-Ven1}.
It is worth noting that not only isotropic Friedmann models
but also anisotropic Bianchi-I models~\cite{Copeland,Copeland1} were studied in this framework.
In these works, the universe contained a dilaton supplemented by an antisymmetric tensor field,
which influenced its dynamics.
Another approach to the problem of the singularity, also inspired by superstring theories, was developed
in Refs.~\cite{Khoury,Khoury1,Khoury2}.
In Ref.~\cite{Khoury1}, the authors treated the singularity as the transition between
a contracting Big Crunch phase and an expanding Big Bang phase.
A crucial role in their analysis was played by a massless scalar field, a modulus.
The theory was reformulated so as to employ variables that remain finite as the scale factor shrinks to zero,
which suggests a natural way to match the solutions before and after the singularity.
The general features of the approach in Refs.~\cite{Khoury,Khoury1,Khoury2}
are the role played by the scalar field and the construction of variables which are finite 
at the singularity crossing.
These features are also essential for the approach developed in Refs.~\cite{KPTVV-sing,we-Bianchi,we-Bianchi1,we-KS}. 
\par
Recently, other approaches to the problem of the description of such a crossing
were elaborated in Refs.~\cite{Bars,Bars1,Wetterich,Wetterich1,Prester,Prester1,Mercati,Mercati1}.
Behind these approaches are basically two general ideas.
Firstly, to cross the singularity, one must give a prescription matching non-singular,
finite quantities before and after the crossing.
Secondly, such a description can be achieved by using a convenient choice of
field parametrizations. 
In Ref.~\cite{KPTVV-sing}, a version of the description of the crossing of singularities
in universes filled with scalar fields was elaborated based on the transition between the Jordan and the Einstein frames.
The main idea of Ref.~\cite{KPTVV-sing} was the following:
when in the Einstein frame the universe arrives at the Big Bang{--}Big Crunch singularity, from the point of view of
the evolution of its counterpart in the Jordan frame, its geometry is regular, but the effective Planck mass is zero.
The solution to the equations of motion in the Jordan frame is smooth at this point,
and using the relations between the solutions of the cosmological equations
in the two frames, one can describe the crossing of the cosmological singularity
in a uniquely determined way.
In Ref.~\cite{Starobinsky1981}, it was pointed out that in a homogeneous and isotropic universe,
one can indeed cross the point where the effective gravitational constant changes sign. 
However, the presence of anisotropies or inhomogeneities changes the situation
drastically because these anisotropies and inhomogeneities grow indefinitely
when the value of the effective Planck mass tends to zero,
hence the effective gravitational constant diverges.
In Ref.~\cite{we-Bianchi}, the authors investigated this phenomenon, suggesting a simple field
reparametrization that allows one to describe the Big Bang{--}Big Crunch singularity crossing
in the Bianchi-I model filled with a minimally coupled scalar field.
Further studies of the description of the Big Bang--Big Crunch singularity
crossing in anisotropic models were undertaken in Refs.~\cite{we-Bianchi1,we-KS}. 
An attempt to develop a general approach to the problem of the treatment
of the singularities in classical and quantum cosmology based on the
reparmetrizations in field space was also undertaken in Refs.~\cite{Ibere,Ibere1,Ibere2}.   
\par
Taking into account all of the above, we shall try to describe what happens with
our solution when the Bianchi-I universe with a magnetic field encounters
the singularity. 
Let us  tackle the problem by studying the behaviour of the differential equations ``behind the singularity'',
where these equations and their solutions remain mathematically well defined but variables can take on
a different meaning from their original one.
For example, the variable $V(t)$, defined as a volume, can become negative.
Let us come back to Eq.~\eqref{magn33}, describing the contraction of the universe when the parameter $\alpha_0$ is positive. 
We have seen that such a universe reaches $V(t_1)=0$ when the velocity of the contraction is
$\dot V(t_1)=-W_1=-2\alpha_0+\sqrt{\alpha_0^2-\beta_0^2}$.
If we continue the solution for $t > t_1$, the variable $V$ becomes negative,
its time derivative remains negative, and it does not satisfy the inequality~\eqref{magn41}.
It follows from Eq.~\eqref{magn33} that the second derivative $\ddot{V}$ remains positive.
Thus, $V(t)$ undergoes some kind of ``negative expansion'' for a period of time
during which it remains negative while its absolute value grows.
Since $\ddot{V}>0$, at some moment, say $t=t_3$, the time derivative passes
through the value $\dot{V}(t_3)=0$ and becomes positive.
Correspondingly, the function  $V(t)$ also starts to increase (its absolute value is decreasing,
so we can speak about ``negative contraction'').
At a later moment, $t=t_4$, the spatial  volume $V(t)$ should vanish.
However, the velocity $\dot{V}>0$ and we cannot enter the interval~\eqref{magn41}
if the parameter $\alpha_0$ is positive, hence the critical $-W_1=-2\alpha_0+\sqrt{\alpha_0^2-\beta_0^2}$ is negative.
This implies that we encounter a singularity of another type when $V(t) \sim \sqrt{t_4-t}$, which we have described before.
And it is impossible to cross this singularity, coming back to the ``normal'' part of the phase space $(V,\dot{V})$
of our problem, where we were before the first encounter with the singularity and where the magnetic field was well defined. 
\par
Nevertheless, a nice way to come back to the ``normal universe'' after this journey behind the ``looking glass'' still exists.
It consists of a simple prescription: we can change the sign of the constant $\alpha_0$.
The reasonable moment for this change is exactly at $t=t_3$, because the parameter
$\alpha_0$ enters Eq.~\eqref{magn33} either squared or multiplied by $\dot{V}$
and the velocity $\dot{V}(t_3)=0$.
To keep the parameters $\alpha_0$ and $\beta_0$ on equal footing, we can change
the sign of the parameter $\beta_0$ as well. 
After that, the universe goes towards the singularity, where $V(t_4) = 0$,
$\dot{V}(t_4)=W_3=-2\alpha_0-\sqrt{\alpha_0^2-\beta_0^2} > 0$, and crosses into
the ``normal'' part of the phase space expanding. 
\par
Now, it is interesting to compare the anisotropy parameters of the universe before
it hits the singularity and after it jumps out of it.
Using the above solution $\sigma = W_1$ to Eq.~\eqref{dif9},  
we found the expressions~\eqref{dif19}-\eqref{dif20} for the scale factors $a(t)$, $b(t)$, and $c(t)$
when the universe is contracting towards the singularity at $t=t_2$.
Let us choose
$\beta_0 =\xi \, \alpha_0$ and $0 \leq \xi \leq 3/5$.
In this case, $p_3 > p_2$ and 
\begin{equation}
u = \frac{\sqrt{\alpha_0^2-\beta_0^2}}{\alpha_0+\beta_0-\sqrt{\alpha_0^2-\beta_0^2}},
\label{dif55}
\end{equation}
so $\xi$ is just like in Eq.~\eqref{dif39}. 
Simple calculations then show that the Kasner indices of the universe that go out from
the region behind the singularity coincide with those it had before the
encounter with the singularity.
It is not clear how we can interpret the period of time that the universe has spent behind the singularity. 
\par
One can suggest another way to describe the transition through the singularity.
The Einstein equations for our system are invariant under a change in the sign
of the scale factors.
Indeed, these equations only contain terms like ${\ddot{a}}/{a}$, 
${\dot{a}}/{a}$ or $a^2$.
Thus, we can change the signs of all three scale factors so that the volume becomes positive again.
To make the equation~\eqref{magn33} for the volume invariant with respect to
the change of sign of $V(t)$, we should also change the sign of the parameter $\alpha_0$.
In this way, the positive time derivative of the volume $\dot{V}$ immediately acquires
a lower critical value $2|\alpha_0| - \sqrt{\alpha_0^2-\beta_0^2}$ and an unlimited expansion begins.  
\par
Both prescriptions described above imply the same evolution after the singular
bounce, but these two bounces have a slightly different origin.
\par
Finally, we would like to make a comment concerning the behavior of the magnetic field as the
universe tends toward the singularity.
Naturally, the energy density of the magnetic field diverges when the universe enters a Kasner regime
approaching the singularity. 
Namely, this density behaves as $ a^{-2}b^{-2} \sim t^{-2(1-p_3)}$.
Nevertheless, if the Kasner index $p_3$ is close to 1, this divergence is much milder than
the one in a isotropic universe.
In any case, as we have explained before, the presence of the diverging quantities does not prevent
us from describing the singularity crossing. 
\section{Concluding remarks}
\label{s:conc}
We have studied in detail some features of the Bianchi-I universe filled with a spatially homogeneous
magnetic field.
Our finding is that the universe is in the Kasner regime at the beginning and end of cosmological evolution.
We have established the relationship between the parameters of these two regimes and shown that
it coincides with the one in the empty Bianchi-II universe, which goes out of the initial Kasner regime
rapidly and enters the final Kasner regime earlier than the Bianchi-I universe filled with a magnetic field.
In addition to a spatially homogeneous magnetic field, the universe also containing dust undergoes
the process of isotropization, while the presence of a massless scalar field implies a modification
of the relations between Kasner indices in two asymptotic regimes.
\par
Here we would also like to briefly mention other recent works in which different aspects of the Bianchi-I
geometry were studied.
First of all, let us note that the role of the time coordinate and of one of the spatial coordinates
in the expression for the metric~\eqref{Bianchi-I} can be
exchanged~\cite{Land-Lif}.
This possibility was already considered in the earliest works~\cite{Kasner,Weyl,Levi-Civita}
and the study of the spatial Kasner-like geometries (in particular, in the presence of thin or thick walls or slabs) still attracts researchers~\cite{Amundsen,Fulling0,Fulling,Kam-Vard,Kam-Vard1}.   
The geodesics in Kasner's universe were considered in Ref.~\cite{Harvey}.
Applications of Kasner-type spacetimes to cosmic jets were done in Refs.~\cite{Mashhoon,Mashhoon1},
and the behavior of fields and particles with spin in Bianchi-I universes was studied~\cite{Saha,Kam-Ter}.
In recent papers~\cite{Parnovsky,Parnovsky1}, the comparison of the Bianchi-I geometry
with the real very early universe and our late-time universe was undertaken,
while Ref.~\cite{Horn} was devoted to the study of Horndeski theory and Ref.~\cite{Chiou:2007sp}
to loop quantum cosmology in the Bianchi-I universe.
Thus, consideration of the simplest non-isotropic cosmological models still flourishes and can bring surprises.
\acknowledgements
R.C., A.K.~and P.P.~are partially supported by the INFN grant FLAG.
The work of R.C.~has also been carried out in the framework of activities of the
National Group of Mathematical Physics (GNFM, INdAM).
The work of P.M. was supported by the Hellenic Foundation for Research
and Innovation (H.F.R.I.), under the ‘Third Call for H.F.R.I. PhD Fellowships’
(Fellowship No.~74191).

\end{document}